\documentclass[prl, fleqn, twocolumn, showpacs]{revtex4}

\usepackage[dvips]{epsfig}

\begin{document}

\title{Orbital Glass in FeCr$_2$S$_4$}
\author{V. Tsurkan$^{a,b}$, V. Fritsch$^{a}$, J. Hemberger$^{a}$,
A. Krimmel$^{a}$, M. M\"ucksch$^{a}$, N. B\"uttgen$^{a}$, H.-A.
Krug von Nidda$^{a}$, D. Samusi$^{b}$,  S. K\"orner$^{a}$, E.-W.
Scheidt$^{a}$, M. Honal$^{a}$, S. Horn$^{a}$, R. Tidecks$^{a}$,
and A. Loidl$^{a}$}

\affiliation{$^{a}$ Institut f\"ur Physik, Universit\"at Augsburg,
D- 86159 Augsburg, Germany}

\affiliation{$^{b}$ Institute of Applied Physics, Academy of
Sciences of Moldova, MD-2028, Chisinau, R. Moldova}




\email{Vladimir.Tsurkan@physik.uni-augsburg.de}

\date{\today}

\begin{abstract}
Low-temperature heat-capacity investigations on the spinel
FeCr$_2$S$_4$ with ferrimagnetic spin order and orbitally
degenerated Jahn-Teller active Fe$^{2+}$ ions in a tetrahedral
crystal field, provide experimental evidence of an orbital liquid
state above 10 K. We demonstrate that the low-temperature
transition at 10 K arises from orbital order and is very
sensitive to fine tuning of the stoichiometry in polycrystals. In
single crystals the orbital order is fully suppressed resulting
in an orbital glass state with the heat capacity following a
strict $T^2$ dependence as temperature approaches zero.
\end{abstract}

%


\maketitle

Orbital physics has become a fascinating topic in modern
solid-state physics. The spatial orientation of orbitals governs
the magnetic exchange and thus the long-range order of the spin
degrees of freedom. If the orientational order of the orbitals
can be changed by an external field (strain field or electric
field), the magnetic order will be changed concomitantly. Such a
tuning of electronic orbitals will be an important ingredient of a
future correlated-electron technology \cite{Tokura2000}.

Strong coupling of spin and orbital degrees of freedom yields
complex ground states. Usually electron-phonon coupling lifts the
orbital degeneracy and results in a long-range orbital order (OO)
and in a change of the crystal symmetry via the Jahn-Teller (JT)
effect \cite{Jahn1937}. In Mott-Hubbard (MH) insulators the OO
may be also established by the Kugel-Khomskii mechanism
\cite{Kugel1982}, where the orbital degeneracy is removed via a
purely electronic interaction. However, recently it has been
suggested that the exchange in orbitally degenerate systems is
strongly frustrated and in cubic lattices the orbitals may remain
disordered down to zero temperature, forming an orbital liquid
(OL) \cite{Ishihara1997,Feiner1997}. Indeed, an OL state has been
proposed for LaTiO$_3$ \cite{Khaliullin2001,Keimer2000}, but has
been questioned recently \cite{Hemberger2003,Cwik2003}. To find
new OL's, it seems promising to search for systems with strong
correlation effects and weak JT coupling. The appropriate
candidates are MH insulators with electrons with threefold
degeneracy in octahedral or twofold degeneracy in tetrahedral
crystal fields \cite{Khomskii2003}, and where in addition the
orbitals are coupled via strongly frustrated exchange
interactions induced by disorder or geometric frustration.
Frustration results in a dynamic liquid ground state or in a
glassy freezing of the internal degrees of freedom at low
temperatures. As well established by numerous studies,
frustration yields a spin-glass state for structurally disordered
spin systems \cite{Binder1986}, while in geometrically
spin-frustrated \cite{Ramirez2001} compounds it results in a
low-temperature spin-liquid \cite{Canals1998} or spin-ice
\cite{Ramirez1999,Bramwell2001} state. In contrast, reports on OL
or JT-glasses, i.e. orbital glasses (OG), are rare
\cite{Ivanov1983,Mehran1983,Babinskii1993}, although OO has been
observed in a variety of $d$-electron systems.

Here we provide experimental evidence for an orbital liquid and
an orbital glass state in a ferrimagnet with perfect spin order.
We focus on the orbital degrees of freedom of the magnetic
semiconductor FeCr$_2$S$_4$ which recently attracted considerable
attention due to its colossal magnetoresistance effect
\cite{Ramirez1997}. It crystallizes in a normal spinel structure
$AB_2$S$_4$ (space group $Fd\bar{3}m$) in which S$^{2-}$-anions
build a cubic close-packed $fcc$ lattice, Fe$^{2+}$-cations
occupy 1/8 of the tetrahedrally coordinated $A$-sites, and
Cr$^{3+}$-cations occupy 1/2 of the octahedrally coordinated
$B$-sites. The Cr$^{3+}$ sublattice (electronic configuration
$3d^3$, spin $S = 3/2$), is dominated by ferromagnetic exchange
of the 90$^{\circ}$ Cr-S-Cr bond angle. The Fe$^{2+}$ ions
($3d^6$, $S = 2$) are only weakly coupled within the sublattice,
but much stronger to the Cr ions \cite{Gibart1969}. Below the
Curie temperature, FeCr$_2$S$_4$ reveals a ferrimagnetic spin
order with the Fe$^{2+}$ and Cr$^{3+}$ magnetic moments aligned
antiparallel to each other \cite{Shirane1964}. Concerning the
orbital degrees of freedom, the orbital moment of the Cr$^{3+}$
ion is quenched. The lower orbitally degenerated $e$-doublet of
Fe$^{2+}$ is JT active allowing for a distortion of the ideal
FeS$_4$ tetrahedron in order to lower the symmetry and, hence to
lift the degeneracy \cite{Feiner1982}. Indeed, an abrupt change
of the electric field gradient detected in M\"ossbauer
experiments \cite{Spender1972} on polycrystalline FeCr$_2$S$_4$
at 10 K was attributed to a transition from a dynamic into a
static JT distortion below 10 K. Additionally, a $\lambda$-type
anomaly of the specific heat $C_p$ at 9.25 K \cite{Lotgering1975}
was registered on Fe-deficient samples. More detailed M\"ossbauer
studies \cite{Brossard1979} have shown that the low-temperature
anomaly cannot be explained in a single-ion picture alone but
involves hybridization of the Fe$^{2+}$ ground state with the
nearest neighbour Cr$^{2+}$ excited state. These results were
interpreted in terms of a transition from orbital paramagnetism
into orbital ferromagnetic order. Such an OO was also proposed
based on band-structure calculations \cite{Park1999}. Recent
ultrasonic experiments on FeCr$_2$S$_4$ single crystals revealed
an elastic anomaly at 60~K, followed by a softening of the
elastic moduli at lower temperature. Below 15 K an additional
anomaly in the longitudinal modulus was observed
\cite{Maurer2003}. These results also suggest a certain kind of
OO.

The present paper focuses on the heat-capacity of FeCr$_2$S$_4$
single crystals and polycrystals with different stoichiometry
tuned by heat treatments in vacuum and sulphur atmosphere.
Polycrystals were prepared by solid-state reactions at
700$^{\circ}$C from the high-purity elements. The as-grown
polycrystals were obtained via repeated homogenisation sintering
(4-5 times) followed by subsequent treatment in sulphur
atmosphere (denoted as $PC_{\rm AG}$). Some were annealed in
vacuum at 700$^{\circ}$C for two hours ($PC_{\rm VA}$) and
subsequently in sulphur atmosphere ($PC_{\rm SA}$) at
700$^{\circ}$C for 24 hours. Single crystals ($SC$) were grown by
chemical-transport reaction at 820 - 850$^{\circ}$C. Rietveld
refinement of x-ray (CuK$_{\alpha}$) powder-diffraction patterns
confirmed the spinel structure and yielded a lattice constant
$a_0 = 1.0005(1)$~nm and a sulphur fractional coordinate $x =
0.259(1)$ for the powdered single crystals. The respective values
of $a_0= 0.9997(1)$~nm for the polycrystals were found to
coincide within the accuracy of the experiment. From a
single-crystal x-ray diffraction analysis of 96 independent and
altogether 2614 Bragg peaks, we determined the site occupancy of
Fe:Cr = 1:1.99 and of Cr:S = 1:2.07, which reveals that the Fe:Cr
ratio is almost ideal while Fe and Cr are slightly ($\sim$ 3\%)
under-stoichiometric when compared to the ideal sulphur network.
An ICP (inductively coupled plasma) analysis of $PC_{\rm AG}$
yielded a nearly stoichiometric Fe:Cr:S ratio of 0.99:1.99:4.0.
The magnetization was measured in a commercial SQUID magnetometer
MPMS-7 (Quantum Design). The heat capacity was investigated in
$^4$He cryostats using adiabatic and relaxational techniques for
$2 \leq T \leq 50$~K and in a $^3$He/$^4$He dilution refrigerator
using relaxational technique for $0.05 \leq T \leq 2.5$~K.

\begin{figure}
\includegraphics[width=6cm]{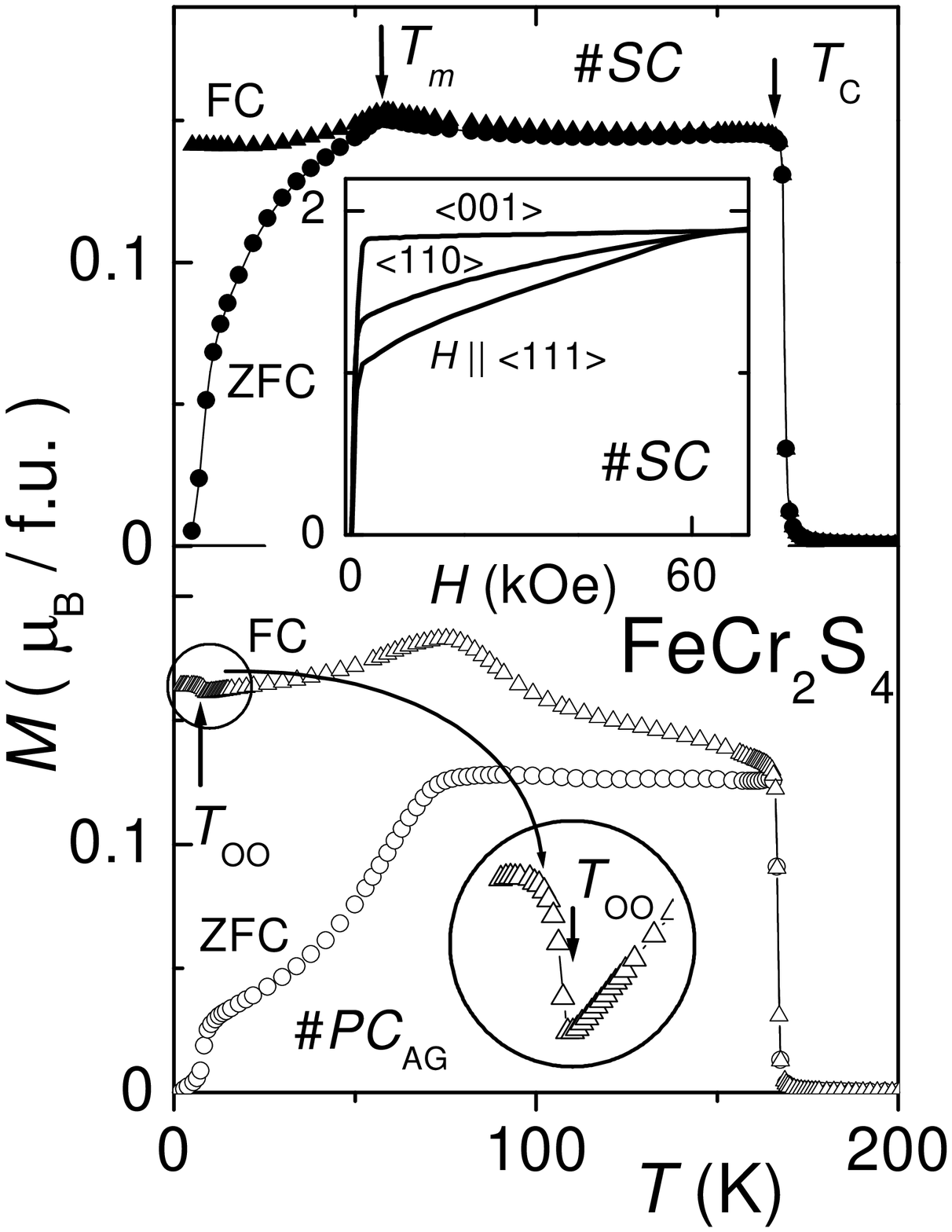}
\caption{Temperature dependence of the zero-field cooled (ZFC)
and field cooled (FC) magnetization $M$ in a field of $H = 100$~Oe
for a FeCr$_2$S$_4$ single crystal ($SC$, upper frame) and in a
field of 50~Oe for an as-grown polycrystal ($PC_{\rm AG}$, lower
frame). Inset: $M(H)$-dependence for $H$ applied along easy
$<001>$, intermediate $<110>$, and hard $<111>$ axes at 4.2~K for
$SC$. \label{mag}}
\end{figure}

Figure~\ref{mag} introduces the basic magnetic properties of
FeCr$_2$S$_4$. It shows the temperature dependence of
magnetization $M(T)$ for single $(SC)$ and polycrystalline
as-grown ($PC_{\rm AG}$) samples at low fields. At the Curie
temperature ($T_{\rm C} = 167$~K) the step-like anomaly in $M(T)$
reveals the onset of ferrimagnetic order. In the single crystal, a
cusp at $T_m \approx 60$~K and a subsequent splitting of the
field cooled (FC) and zero-field cooled (ZFC) magnetization
indicate domain-reorientation processes \cite{Tsurkan2001}. While
in the polycrystal the overall magnetization behavior is similar,
distinct differences can be observed: here the ZFC and FC curves
split already at $T_{\rm C}$ due to pinning of the magnetic domain
walls by grain boundaries. The cusp-like anomaly is also present
in the polycrystals, although shifted to higher $T$ ($\approx
75$~K). Below this anomaly both the poly- and single crystals
exhibit a pronounced downturn in the ZFC magnetization. However,
only the polycrystal shows a clear step-like decrease (increase)
in the ZFC (FC) magnetization at $T_{\rm OO} \approx 10$~K, which
we ascribe to the OO transition. The accompanying magnetic
anomaly at $T_{\rm OO}$ coincides with the low-temperature
transition detected earlier by M\"ossbauer experiments in
polycrystals \cite{Spender1972,Brossard1979} associated with a
JT-like transition resulting in long-range OO. The inset of
Fig.~\ref{mag} shows the field dependence of the magnetization
$M(H)$ at $T = 4.2$~K for the FeCr$_2$S$_4$ single crystal. The
saturation magnetization $M_{\rm s}$ of $1.88 \mu_{\rm B}$ per
formula unit (f.u.) is consistent with the ferrimagnetic order of
Cr with $M_{\rm s}(2 {\rm Cr}) = 6 \mu_{\rm B}/{\rm f.u.}$ and Fe
with $M_{\rm s}({\rm Fe}) = 4.12 \mu_{\rm B}/{\rm f.u.}$, taking
into account the slightly enhanced $g$-value of Fe$^{2+}$ $(g
\approx 2.06)$ due to spin-orbit coupling. Thus, a fully
developed spin order below $T_{\rm C}$ is established, as was
already shown by neutron-scattering studies \cite{Shirane1964}.
Thus, the low-temperature anomalies are governed not by spin but
by orbital degrees of freedom.

\begin{figure}
\includegraphics[width=6cm]{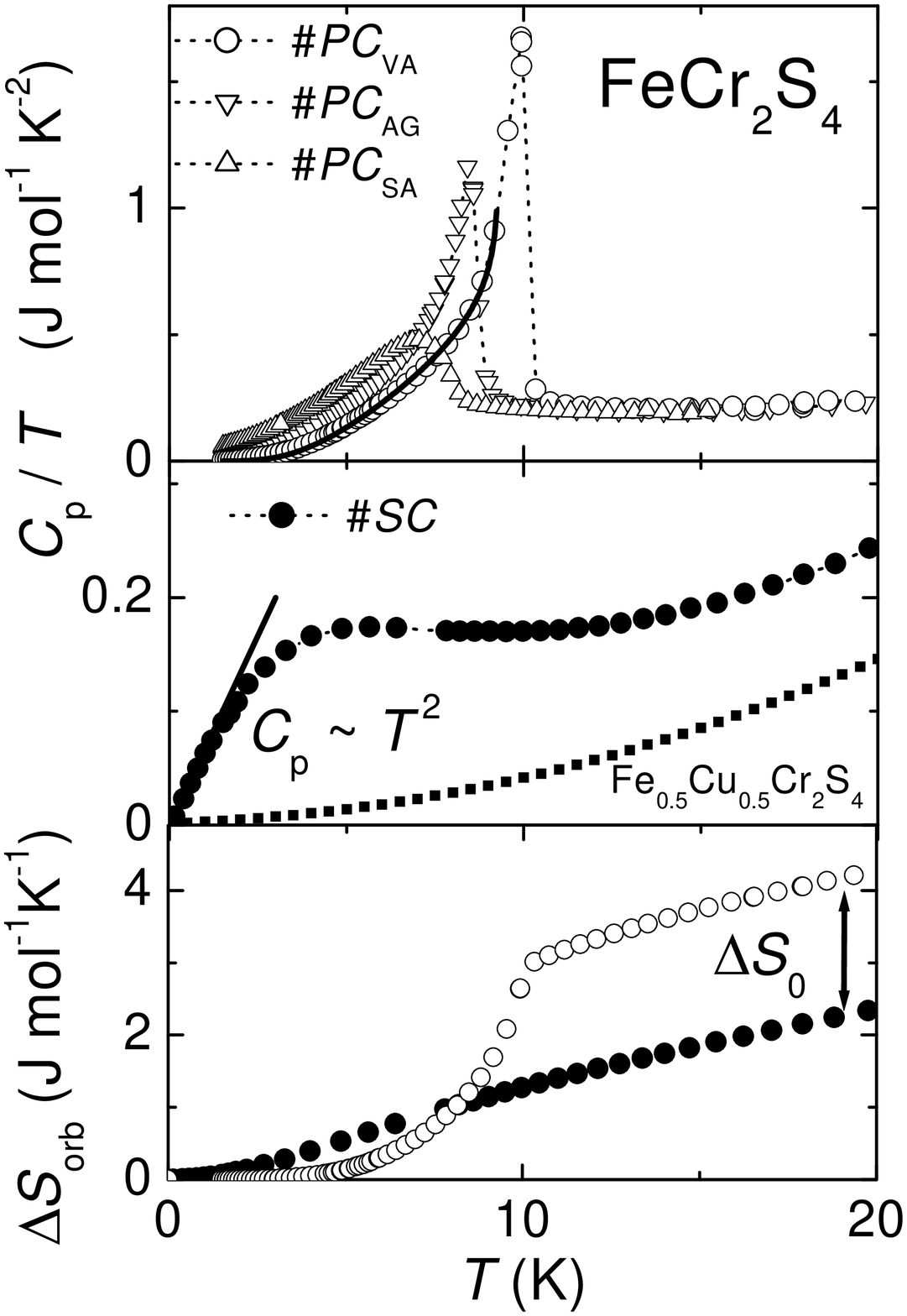}
\caption{Upper frame: $T$-dependence of $C_p/T$ for polycrystals
(open symbols) with different heat treatments: vacuum annealed
($PC_{\rm VA}$), as-grown ($PC_{\rm AG}$), and sulphur-annealed
($PC_{\rm SA}$). Solid line: mean-field fit with $T_{\rm OO} =
9.2$~K and $\Delta (T=0)/k_{\rm B} = 11.4$~K. Middle frame:
$C_p/T$ of the single crystal ($SC$). Solid line indicates $C_p
\propto T^2$ dependence. Solid squares: $C_p/T$ of
Fe$_{0.5}$Cu$_{0.5}$Cr$_2$S$_4$. Lower frame: orbital entropy
change $\Delta S_{\rm orb}(T)$ for the $SC$ and the $PC_{\rm VA}$
calculated as described in the text.\label{cp}}
\end{figure}

Figure~\ref{cp} illustrates the principal results of the present
investigation. The upper frame shows the specific heat at low $T$
with axes $C_p/T$ {\it vs.} $T$ as measured in three
polycrystalline samples after different heat treatment. The
middle frame presents the specific heat of the single crystal.
The vacuum annealed $PC_{\rm VA}$ and as-grown $PC_{\rm AG}$
polycrystals reveal well-defined OO transitions. In the
sulphur-annealed $PC_{\rm SA}$ the phase-transition temperature
is strongly reduced and the transition is smeared out. Moreover,
the transition is completely suppressed in the single crystal. In
the polycrystals the $\lambda$-type anomaly is followed by an
exponential decrease of $C_p$ with decreasing $T$, indicative of
a gap opening in the orbital excitations. For the polycrystal
$PC_{\rm VA}$ with the strongest $C_p$ anomaly the data can be
described within a BCS-like mean-field approach using a
$T$-dependent gap energy $\Delta (T)$ approximated by $\Delta (T)
\approx 1.74\Delta _0(1-T/T_{\rm OO})^{1/2}$. The best fit is
obtained with the parameters $\Delta _0/k_{\rm B} = 11.4$~K and
$T_{\rm OO} = 9.2$~K, yielding $2\Delta _0/k_{\rm B} \approx 2.5
T_{\rm OO}$, which is reduced as compared to the BCS value,
suggesting a weak coupling scenario.

On decreasing $T$, the specific heat in the single crystal passes
through a cusp-shaped maximum and then approaches zero following
a strict $T^2$-dependence. In canonical spin glasses a cusp-like
maximum is also observed slightly above the freezing temperature,
but a linear term evolves approaching $T=0$ \cite{Binder1986}. The
$T^2$-dependence of the specific heat below the maximum is
theoretically not expected for three-dimensional disordered
spin-systems, but has also been observed in the geometrically
frustrated spinel AlV$_2$O$_4$ \cite{Matsuno2003} as well as in
geometrically frustrated two-dimensional spin glasses
\cite{Ramirez1990}. Ac-susceptibility measurements in
FeCr$_2$S$_4$ single crystals ruled out a spin-glass state
\cite{Tsurkan2001}. Thus, the absence of the $\lambda$-type
anomaly and the cusp-shaped maximum observed in FeCr$_2$S$_4$
single crystals suggest a glassy-like state of the orbitals.

For $T > T_{\rm OO}$ all samples reveal a very similar specific
heat with a highly enhanced linear term. This, however, cannot be
explained by phonon or magnon contributions, which were estimated
by measuring the specific heat of the related compound
Fe$_{0.5}$Cu$_{0.5}$Cr$_2$S$_4$, which has similar atomic mass and
Debye temperature as FeCr$_2$S$_4$. In this compound the Cu ions
are monovalent and all Fe ions have a valence of $3+$ ($3d^5$, $S
= 5/2$) and hence are not JT active. It is a simple ferrimagnet
whose specific heat can be well described by superposition of a
$T^3$-law for phonons and a $T^{3/2}$-law for magnons. Comparison
with Fe$_{0.5}$Cu$_{0.5}$Cr$_2$S$_4$ thus reveals an additional
linear contribution to the specific heat $C_p$ of the order of
$\Delta C_p/T \approx 100$~mJ/molK$^2$ in FeCr$_2$S$_4$. It
cannot be attributed to an enhanced Sommerfeld coefficient,
because FeCr$_2$S$_4$ is insulating at low $T$
\cite{Ramirez1997,Park1999}. Therefore, we ascribe it to strongly
fluctuating orbitals. It is noteworthy that a linear term in the
$T$-dependence of the heat capacity has been already predicted
for the OL \cite{Khaliullin2000}. Furthermore, the existence of a
dynamic JT effect had been originally deduced from M\"ossbauer
experiments \cite{Spender1972} and later was termed orbital
paramagnetism regarding its cooperative properties involving the
$B$-site Cr states \cite{Brossard1979}. The dynamic JT effect as
well as orbital paramagnetism indicate orbital fluctuations and
in their cooperative manner provide an OL state.

So far, we conclude that for 10~K~$< T < T_m$ in FeCr$_2$S$_4$ an
OL state is formed with fluctuating orbital arrangement, the
average of which is probably detected as a softening of elastic
constants below 60 K by ultrasound experiments \cite{Maurer2003}.
The gap predicted for an OL in the orbital excitation spectrum
\cite{Khaliullin2001} must be very small, as in the single crystal
the linear term persists down to approximately 3~K and no
exponential decay is observed at lower $T$. In polycrystals a
transition occurs from the OL into the OO state at 10~K. In the
single crystal the OO transition is suppressed and, taking into
account the cusp-like maximum and decrease of the specific heat
at lowest $T$, we suggest that the orbital degrees of freedom
undergo a freezing transition into an OG state with randomly
frozen-in orbital configurations.

\begin{figure}
\includegraphics[width=6.0cm]{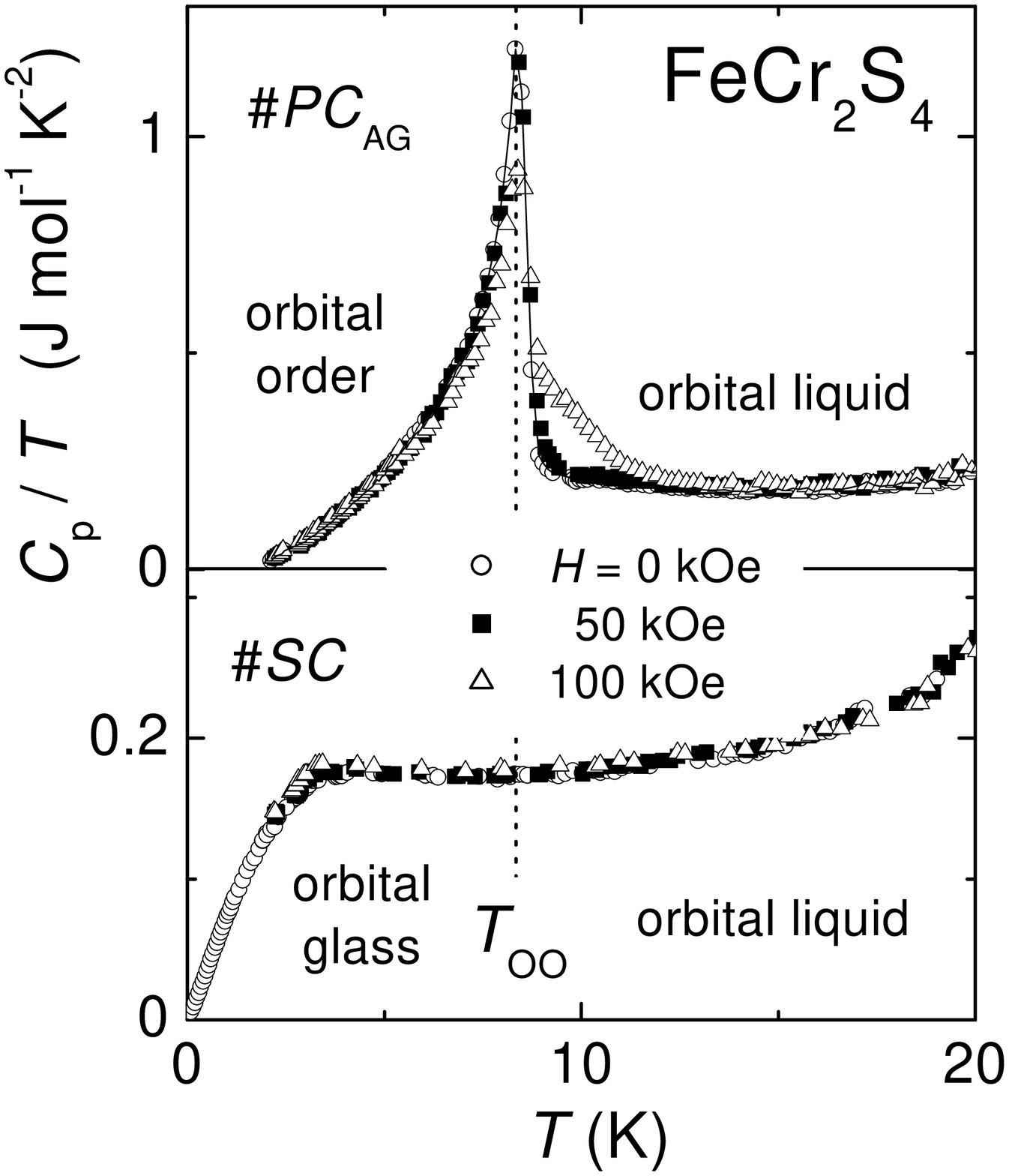}
\caption{$T$ dependence of $C_p/T$ for the as-grown polycrystal
($PC_{\rm AG}$, upper frame) and single crystal ($SC$, lower
frame) at different magnetic fields. Dash line: $T_{\rm OO}$ for
$PC_{\rm AG}$.\label{cph}}
\end{figure}

The heat-capacity experiments in different external magnetic
fields provide further support that the order-disorder transition
in the polycrystal and the glassy freezing in the single crystal
originate from the orbital degrees of freedom. Figure~\ref{cph}
illustrates the effect of an external magnetic field on the
specific heat. The magnitude of the $\lambda$-anomaly in the
polycrystal becomes slightly reduced, but without any noticeable
shift in the ordering temperature. An increase of the specific
heat becomes apparent just above $T_{\rm OO}$ only at 100~kOe. In
the single crystal the specific heat is essentially field
independent. If the $\lambda$-anomaly was related to a transition
in the spin system, a much stronger influence of the magnetic
field is expected, as e.g. observed in MnCr$_2$S$_4$
\cite{Tsurkan2003}. Therefore, the $\lambda$-anomaly in the
specific heat of FeCr$_2$S$_4$ should be attributed to the onset
of OO.

Now we turn to the evaluation of entropy for poly- and single
crystals shown in the lowest frame of Figure~\ref{cp}. The change
of orbital entropy $\Delta S_{\rm orb}$ has been obtained by
integrating $\Delta C_p(T)/T$ for $0.1 \leq T \leq 20$~K. In the
OO system ($PC_{\rm VA}$), The orbital entropy $S_{\rm orb}$
should approach zero below the OO transition. Above $T_{\rm OO}$,
it is still increasing, indicating that the total value $R\ln{2}
= 5.8$~J/molK assigned to the orbital doublet has not been fully
recovered. It will be recovered at higher $T$ due to increasing
orbital fluctuations, as expected for an OL. For the single
crystal the all-over change of orbital entropy below 10~K is
about $\Delta S_0 = 2$~J/molK smaller than for the polycrystal.
This difference of 1/3 of the total orbital entropy clearly
demonstrates a macroscopic degeneracy of the ground state in the
single crystal, which we interpret in terms of an orbital glass
as $T$ approaches zero.

This result surprisingly is quite similar to that observed in
spin-ice systems \cite{Ramirez1999}. Keeping in mind the observed
$T^2$ dependence of the specific heat and the tetrahedral network
of the Fe$^{2+}$ ions, one may assume a certain role of
geometrical frustration in the formation of the OG in the single
crystal. To decide this definitely, a more precise knowledge of
the orbital exchange is necessary and the connection between spin
and orbital order needs further theoretical analysis.

In summary, our heat capacity study of FeCr$_2$S$_4$ reveals a
strong linear contribution $C_p \propto T$ in both poly- and
single crystals suggesting an orbital liquid state. In
polycrystals we observed a transition from this orbital liquid
into an orbitally ordered state, which can be strongly shifted by
fine tuning of the stoichiometry. In single crystals the orbital
order is fully suppressed. We suggest that a transition from the
orbital liquid into an orbital glass occurs, which is supported by
the $T^2$-behavior of $C_p$ on approaching $T=0$. Thus, our
results provide experimental evidence of such an orbital-liquid to
orbital-glass transition in a $3d$-magnetic system.

This work was supported by the BMBF via VDI/EKM, FKZ 13N6917/18
and by DFG within SFB 484 (Augsburg).


\begin{thebibliography}{99}

\bibitem{Tokura2000} Y. Tokura and N. Nagaosa, Science {\bf 228}, 462 (2000).

\bibitem{Jahn1937} H. A. Jahn and E. Teller, Proc. Roy. Soc. (London) A {\bf 161}, 220
(1937).

\bibitem{Kugel1982} K. I. Kugel and D. I.  Khomskii,  Sov. Phys. Usp. {\bf 25}, 231
(1982).

\bibitem{Ishihara1997} S. Ishihara, M. Yamanaka, and N. Nagaosa, Phys. Rev. B {\bf 56}, 686
(1997).

\bibitem{Feiner1997} L. F. Feiner, A. M. Oles, and J. Zaanen,  Phys. Rev. Lett. {\bf 78},
2799 (1997).

\bibitem{Khaliullin2001} G. Khaliullin, Phys. Rev. B {\bf 64}, 212405 (2001).

\bibitem{Keimer2000} B. Keimer {\it et al.}, Phys. Rev. Lett. {\bf 85}, 3946 (2000).

\bibitem{Hemberger2003} J. Hemberger {\it et al.}, Phys. Rev. Lett. {\bf 91}, 066403 (2003).

\bibitem{Cwik2003} M. Cwik {\it et al.}, Phys. Rev. B {\bf 68}, 060401 (2003).

\bibitem{Khomskii2003} D. I. Khomskii and M. V. Mostovoy, J. Phys. A {\bf 36}, 9197 (2003).

\bibitem{Binder1986} K. Binder and A. P. Young, Rev. Mod. Phys. {\bf 58}, 801 (1986).

\bibitem{Ramirez2001} A. P. Ramirez, in {\it Handbook of Magnetic Materials}, edited by K. H. J.
Buschow (Elsevier Science, Amsterdam, 2001), vol. 13,  p. 423.

\bibitem{Canals1998} B. Canals and C. Lacroix, Phys. Rev. Lett. {\bf 80}, 2933 (1998).

\bibitem{Ramirez1999} A. P. Ramirez {\it et al.}, Nature {\bf 399}, 333 (1999).

\bibitem{Bramwell2001} S. T. Bramwell and M. J. P. Gingras,  Science {\bf 294}, 1495 (2001).

\bibitem{Ivanov1983} M. A. Ivanov {\it et al.}, J. Magn. Magn. Mater. {\bf 36}, 26 (1983).

\bibitem{Mehran1983} F. Mehran and K. W. H. Stevens, Phys. Rev.  B {\bf 27}, 2899 (1983).

\bibitem{Babinskii1993} A. V. Babinskii {\it et al.}, JETP Lett. {\bf 57}, 299 (1993).

\bibitem{Ramirez1997} A.P. Ramirez, R.J. Cava, and J. Krajewski, Nature {\bf 386}, 156
(1997).

\bibitem{Gibart1969} P. Gibart, J. L. Dormann, and Y. Pellerin, Phys. Stat. Sol. {\bf 36},
187 (1969).

\bibitem{Shirane1964} G. Shirane and D. E. Cox, J. Appl. Phys. {\bf 35}, 954 (1964).

\bibitem{Feiner1982} L. F. Feiner,  J. Phys. C: Solid State Phys. {\bf 15}, 1515
(1982).

\bibitem{Spender1972} M. R. Spender and A. H. Morrish,  Solid State Commun. {\bf 11}, 1417
(1972).

\bibitem{Lotgering1975} F. K. Lotgering, A. M. van Diepen, and J. F. Olijhoek, Solid State
Commun. {\bf 17}, 1149 (1975).

\bibitem{Brossard1979} L. Brossard {\it et al.}, Phys. Rev. B {\bf 20}, 2933 (1979).

\bibitem{Park1999} M. S. Park {\it et al.}, Phys. Rev. B {\bf 59},
10018 (1999).

\bibitem{Maurer2003} D. Maurer {\it et al.}, J. Appl. Phys.
{\bf 93}, 9173 (2003).

\bibitem{Tsurkan2001} V. Tsurkan {\it et al.}, J. Appl. Phys. {\bf 90}, 4639 (2001).

\bibitem{Matsuno2003} K. Matsuno {\it et al.}, Phys. Rev. Lett. {\bf 90}, 096404 (2003).

\bibitem{Ramirez1990} A. P. Ramirez, G. P. Espinosa, and A. S. Cooper, Phys. Rev. Lett.
{\bf 64}, 2070 (1990).

\bibitem{Khaliullin2000} G. Khaliullin and S. Maekawa, Phys. Rev. Lett. {\bf 85}, 3950 (2000).

\bibitem{Tsurkan2003} V. Tsurkan {\it et al.}, Phys. Rev. B {\bf 68}, 134434 (2003).
\end{thebibliography}
\end{document}